\documentclass[twocolumn,showpacs,prb]{revtex4}
\usepackage{graphicx}%
\usepackage{dcolumn}
\usepackage{amsmath}
\usepackage{latexsym}
\usepackage{longtable}
\begin{document}
\title{Size induced change-over from first to second order magnetic phase transition in $\mathrm{La_{0.67}Ca_{0.33}MnO_3}$ nanoparticles.}
    
\author{Tapati Sarkar\footnote[1]{email:tapatis@bose.res.in} and  A. K. Raychaudhuri\footnote[2]{email:arup@bose res.in}}
\affiliation{DST Unit for NanoSciences, S.N.Bose National Centre for Basic Sciences, Block JD, Sector III, Salt Lake, Kolkata - 700 098, India.}
\date{\today}
\begin{abstract}

\noindent
In this report we show that in the perovskite manganite $\mathrm{La_{1-x}Ca_{x}MnO_3}$ for a fixed $x \approx 0.33$, the magnetic transition changes over from first order to second order on reducing the particle size to nearly few tens of a nanometer. The change-over is brought about only by reducing the {\it size} and with no change in the stoichiometry.  The size reduction to an average size of about $15 nm$ retains the ferromagnetic state albeit with somewhat smaller saturation magnetization and the ferromagnetic transition temperature ($T_{C}$)  is suppressed by a small amount ($\sim 4\%$). The magnetization of the nanoparticles near $T_{C}$ follow the scaling equation $M/|\epsilon|^\beta = f_\pm(H/|\epsilon|^{\gamma+\beta})$, where, $\epsilon = |T-T_C|/T_C$. The   critical exponents, associated with the transition have been obtained  from modified Arrott plots  and they  are found to be  $\beta=0.47\pm 0.01$ and $\gamma=1.06\pm 0.03$. From a plot of $M$ vs $H$ at $T_{C}$ we find the exponent  $\delta=3.10 \pm 0.13$. All the exponents are  close to the mean field values. The change-over of the order of the transition has been  attributed to a lowering of the value of the derivative  $dT_{C}/dP$  due to an increased pressure in the nanoparticles arising due to size reduction. This effect acts in tandem with the  rounding off effect due to random strain in the nanoparticles.
\end{abstract}
\maketitle

\noindent
\section{\bf INTRODUCTION}
$\mathrm{La_{0.67}A_{0.33}MnO_3}$ (A = alkaline earth element) shows a  phase transition at a ferromagnetic Curie temperature $T_C$. While in most  doped manganites (A = Sr, Ba etc.) with wider band width the phase transition is  second order, in narrow band width $\mathrm{La_{0.67}Ca_{0.33}MnO_3}$ the transition has been found to be  first order\cite{ref1,ref2,GhoshLSMO}. The issue of the nature of magnetic phase transition in the system $\mathrm{La_{1-x}Ca_{x}MnO_3}$ (LCMO) as the hole concentration $x$ is changed is a topic of considerable interest and has not been understood yet, although experimentally it has been observed that the transition is first order for $x \approx 0.2-0.4$ and is a continuous one for $x$ outside this regime. For $x=0.4$, in particular, there is a tricritical point at the ferromagnetic $T_{C}$\cite{tricritical}. In the Ca substituted manganites which have a smaller band width, the effect of the electron-lattice coupling is stronger and presumably it can play a very dominant role in deciding the nature of the transition in addition to the contribution arising from the double-exchange (DE) interaction which is the cause of ferromagnetism in these materials.

\noindent
 In this paper, we present an interesting observation that the first order phase transition in $\mathrm{La_{0.67}Ca_{0.33}MnO_3}$ (LCMO, x=0.33) changes to a second order transition when the particle size is brought down to the range of few tens of a nanometer. In this change-over the hole concentration (i.e, $x$) is kept fixed and the change over occurs solely due to factors that arise due to size reduction. 

\noindent
It is important to put the present investigation in the context of the nature of phase transition in the LCMO system. In manganites other than LCMO (like $\mathrm{La_{1-x}Sr_{x}MnO_3}$ (x=0.3)\cite{ref1,GhoshLSMO}, $\mathrm{Nd_{1-x}Pb_{x}MnO_3}$ (x=0.4)\cite{NPMO}) the magnetic transition is a continuous transition. The critical exponents are close to what one would expect from a 3D Heisenberg model with short range interactions. Interestingly, the first order transition observed in LCMO (x=0.33) can be made a continuous transition by substitution of Ca in the A-site by small amount Ba or Sr\cite{ref1,ref2} or substitution in Mn sites by Ga\cite{Ga}. The critical exponents obtained in these cases were also what one would expect for a 3D Heisenberg model. The basic magnetic system in the manganites with DE interaction is thus a simple 3D magnetic system. There are  causes like strong coupling to other modes that makes the transition in LCMO in the special hole concentration region (x=0.2-0.4) a first order transition. It is suggested that the transition in LCMO is a  fluctuation-driven first order transition\cite{fluctuation}. The change-over to a second order (continuous) transition on small substitution by Ba or Sr at Ca site or by Ga at Mn site is generally thought as arising from random disorder that rounds off the first order transition to a second order transition\cite{quenched}. The observation  of a size reduction induced continuous transition is an important observation because the system has the same chemical composition (no change in hole concentration, or dilution of the magnetic interaction) and the only difference is the size of the system. This change-over  happens with no significant change in the value of the transition temperature $T_{C}$. Substitution of Sr and Ba enhances $T_{C}$ while dilution of Mn site by Ga reduces $T_{C}$ significantly. To our knowledge, this particular issue has not been addressed before and no experiments have been reported that specially observe this change-over of the nature of the magnetic phase transition on size reduction. The synthesis, characterization and magnetic measurements are described in details in the subsequent sections.

\noindent
The nature of the magnetic transition can be obtained from the slope of isotherm plots of H/M vs. $M^2$, M being the experimentally observed magnetization and H the magnetic field. This criterion of deciding the nature of the transition is generally referred to as the Banerjee criterion\cite{ref3} and has been used widely to experimentally determine the order of the magnetic phase transition. Briefly, a positive or a negative slope of the experimental $H/M$ vs. $M^2$ curve indicates a second order or first order transition respectively. We have applied this criterion to $\mathrm{La_{0.67}Ca_{0.33}MnO_3}$ samples with two widely different particle sizes and found striking differences in the magnetic behaviour of the two samples near $T_C$ when the particle size is brought down from a few $\mu$m to few tens of a nanometer. 

\noindent
\section{ EXPERIMENTAL}
We have adopted the sol - gel based polymeric precursor (polyol) route to synthesize $\mathrm{La_{0.67}Ca_{0.33}MnO_3}$ (LCMO, x=0.33) nanoparticles with sizes down to $15nm$. This method allows synthesis at a significantly lower sintering temperature compared to the conventional solid state procedure. In this technique the polymer (ethylene glycol in our case) helps in forming a close network of cations from the precursor solution and assists the reaction, enabling phase formation at relatively low temperatures\cite{ref4,JAPPCMO}.  A major challenge encountered in the synthesis of nanocrystalline multicomponent oxides is the poor control of stoichiometry at the nano level. However, our synthesis route ensures homogeneity, phase purity and a good control over stoichiometry, as brought out by X-Ray Diffraction (XRD) and chemical tests. The details of sample synthesis have been given elsewhere. We prepared samples using high purity acetates (procured from Sigma - Aldrich\cite{ref5}). We optimized the process parameters to obtain phase pure $\mathrm{La_{0.67}Ca_{0.33}MnO_3}$ particles with size $\sim15 nm$ (as established from XRD results, using the Williamson Hall plot\cite{ref6} as well as from TEM data). We used pellets of the nanopowders for magnetic measurements. For the nano particle sample, we used the unsintered pellet, which retained the average particle size of the as-prepared powder ($\sim$ 15nm). The particle size of the sample can be controlled by heat treatment of the pellet. For comparison with standard data, we prepared a bulk sample by sintering one of the pellets at 1300\r{ }C for 2 hrs. This caused grain growth, giving an average particle size of a few $\mu$m, for the bulk sample.

\noindent
 Both the samples  were  characterized  using  powder  X-Ray  Diffraction   (XRD)  using  CuK$\alpha$  radiation  at  room temperature. The nanometer sized samples were also characterized for their size and crystallinity by a Transmission Electron Microscope (TEM) and the sample with larger grain size (the bulk sample) was also characterized using a Field Emission Scanning Electron Microscope (FE-SEM). The  pellets  were  also checked  for  oxygen  stoichiometry  using  iodometric  titration.  The  magnetic  measurements have  been  carried  out using a Vibrating Sample Magnetometer\cite{ref7}. Magnetization isotherms at various temperatures around the critical region have been measured in applied fields upto 1.6T with a temperature control better than $\pm 0.05K$.

\noindent
\section{ RESULTS}
\subsection {STRUCTURE}
\begin{figure}[t]
\begin{center}
\includegraphics[width=8cm,height=7cm]{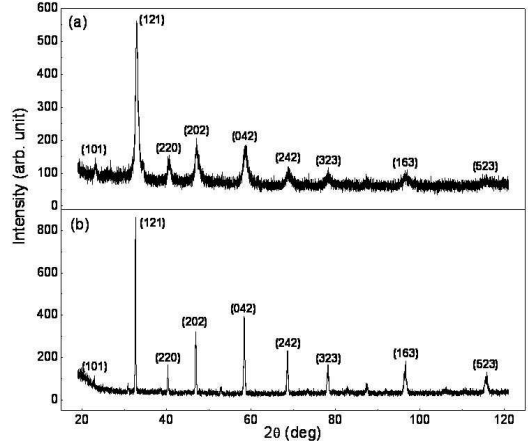}
\end{center}
\caption{XRD pattern of LCMO (a)nanocrystals (average size $\sim 15nm$) and (b)bulk sample (average size $\sim$ 2.7$\mu$m)}
\label{Fig1}
\end{figure}

Fig.~\ref{Fig1} shows the XRD pattern of  LCMO bulk and nanoparticles. The XRD pattern shows pure phase formation in the samples without any impurity phase. The line broadening in the nano particle sample was used to estimate the average particle size of the nanoparticles using the Williamson Hall plot\cite{ref6}. The nanosample and the bulk sample have similar lattice constants inspite of orders of magnitude difference in the size. The $a$ and $b$ lattice constants are smaller  by $\sim 0.5\%$ in the nanoparticle and the $c$ axis is smaller  by $\sim 1.0\%$. The cell volume for the nanoparticle is $\approx 227 A^{3}$ and it is $1.9\%$ smaller than the bulk. We note that the contraction in the cell volume in the nanoparticle can be thought of as arising from an effective hydrostatic pressure of 3-4 GPa (estimated using a bulk modulus of $\approx$ 150-200 GPa\cite{Aveek}). We discuss this issue in details in subsequent sections.

\begin{figure}[t]
\begin{center}
\includegraphics[width=9cm,height=4.2cm]{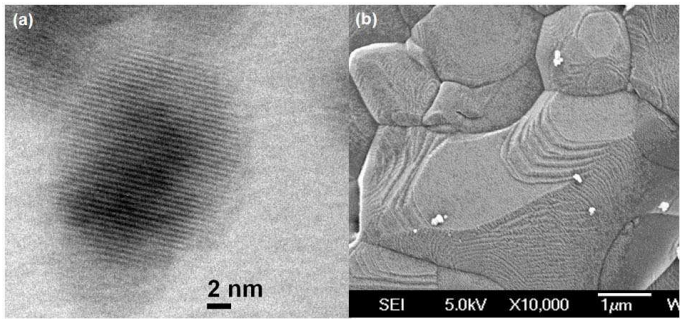}
\end{center}
\caption{(a)HRTEM image of LCMO nanocrystals, (b)FE-SEM image of the bulk sample}
\label{SEMTEM}
\end{figure}

\noindent
Fig.~\ref{SEMTEM}(a) shows the High resolution Transmission Electron Microscopy (HRTEM) image of the nanoparticles. The HRTEM image shows the single crystalline nature of the particles. This is an important observation that the size reduction retains the single crystalline nature of the individual grains. The size distribution as obtained from the TEM image shows that 90\% of the particles are lying within the size range 14nm to 18nm. The mean size as calculated from the TEM image is $\sim 16nm$ with a rms size distribution of $\pm$10{\%}. 
 
In the same Fig.~\ref{SEMTEM}(b) we also show FE-SEM image of the bulk sample. The sample has grains that have sizes well in excess of a few $\mu m$. 90\% of the grains have sizes lying within the range 1.3-3.8$\mu m$ and the average size is 2.7$\mu m$ with a rms size distribution well in excess of that seen for the nanoparticles. We note that though the bulk sample has larger grains it has a somewhat wider size distribution. This is expected as the sample has been prepared by grain growth of the nanosample by heat treatment. The wider size distribution of the bulk sample is worth noting because even in the presence of the size distribution, the sample preserves its first order transition as observed below.  Importantly this method allows us to use chemically the same material as the starting material thus avoiding problems with stoichiometry. The oxygen stoichiometry has been measured  by iodometric titration. For the bulk sample it is 2.97 and for the nano sample it is 2.98. 

\noindent
\subsection{ MAGNETIZATION MEASUREMENTS}
\begin{figure}[t]
\begin{center}
\includegraphics[width=8cm,height=7cm]{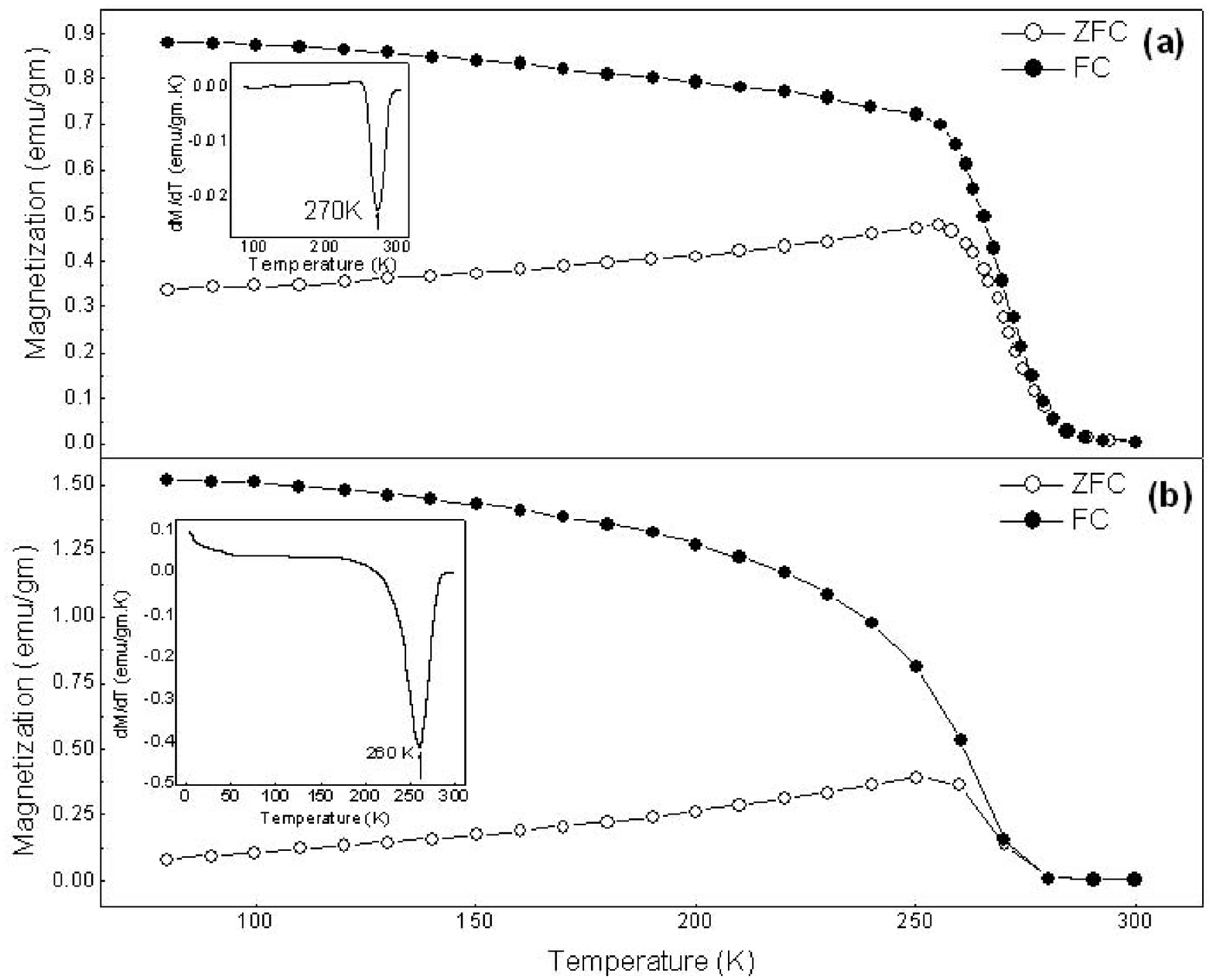}
\end{center}
\caption{Magnetization versus temperature for (a)bulk and (b)nano LCMO under a magnetizing field of 1mT. The insets show dM/dT versus T}
\label{Fig3}
\end{figure}

Low field (H = 1mT) magnetization versus temperature was first measured for both the samples in order to fix  the transition temperatures. The results ($M-T$ curves) for the bulk and the nano particle sample are presented in Fig.~\ref{Fig3}(a) and (b) respectively. The insets in the figures show plots of $dM/dT$ versus $T$. We take the inflection points in the $M-T$ curves to be an estimate of the Curie temperature $T_C$. (The exact value of $T_{C}$  has been obtained by a scaling fit described later on). For the bulk sample this comes out to be 270K. For the nano particle sample $T_C$ is slightly lower ($\sim$ 260K). The transition width ($\Delta T_{1/2}$), roughly defined as the full width at the half maximum for the $dM/dT$ vs T curve, is somewhat larger in the nanopaticles. ($\Delta T_{1/2} \approx 27K$ for the nanoparticle and $\approx 20K$ for the bulk sample).

\begin{figure}[t]
\begin{center}
\includegraphics[width=8cm,height=7cm]{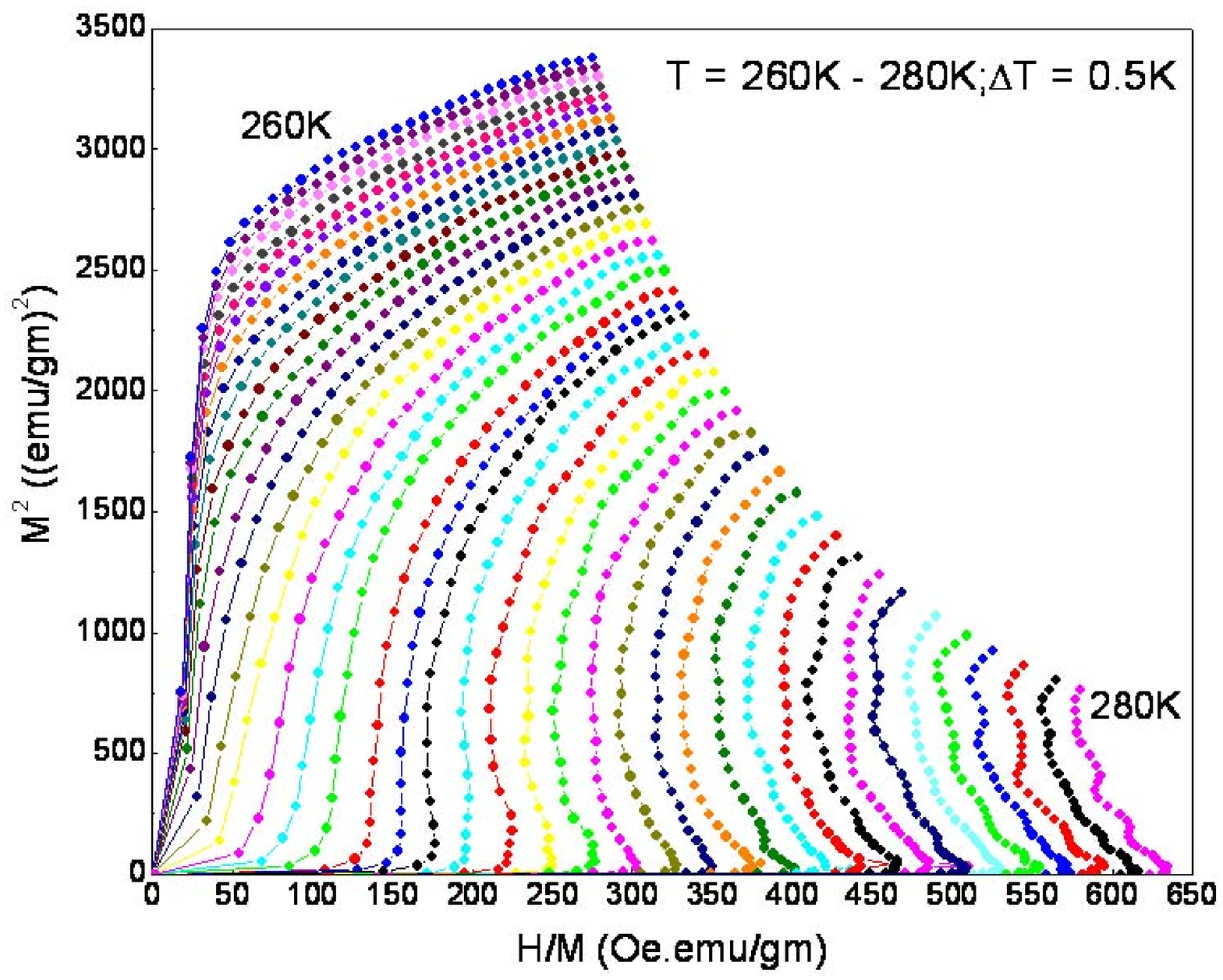}
\end{center}
\caption{Arrott plots for bulk LCMO}
\label{Fig4}
\end{figure}

\noindent
In order to apply the Banerjee criterion and find out the nature of the transition, we have measured initial magnetization isotherms in the vicinity of the critical temperature. Before each run, samples were heated above their $T_C$ and cooled to the measuring temperature under zero field in order to ensure a perfect demagnetization of the samples. In Fig.~\ref{Fig4}, we plot $M^2$ versus $H/M$ isotherms (Arrott plots) between 260K and 280K for the bulk sample. It is clear that the isotherms present negative slopes in some parts which, according to the criterion used here, is an indication of the first order character of the transition. The observation of a first order transition in the bulk sample is in agreement with past studies\cite{ref1,ref2}.

\begin{figure}[t]
\begin{center}
\includegraphics[width=8cm,height=7cm]{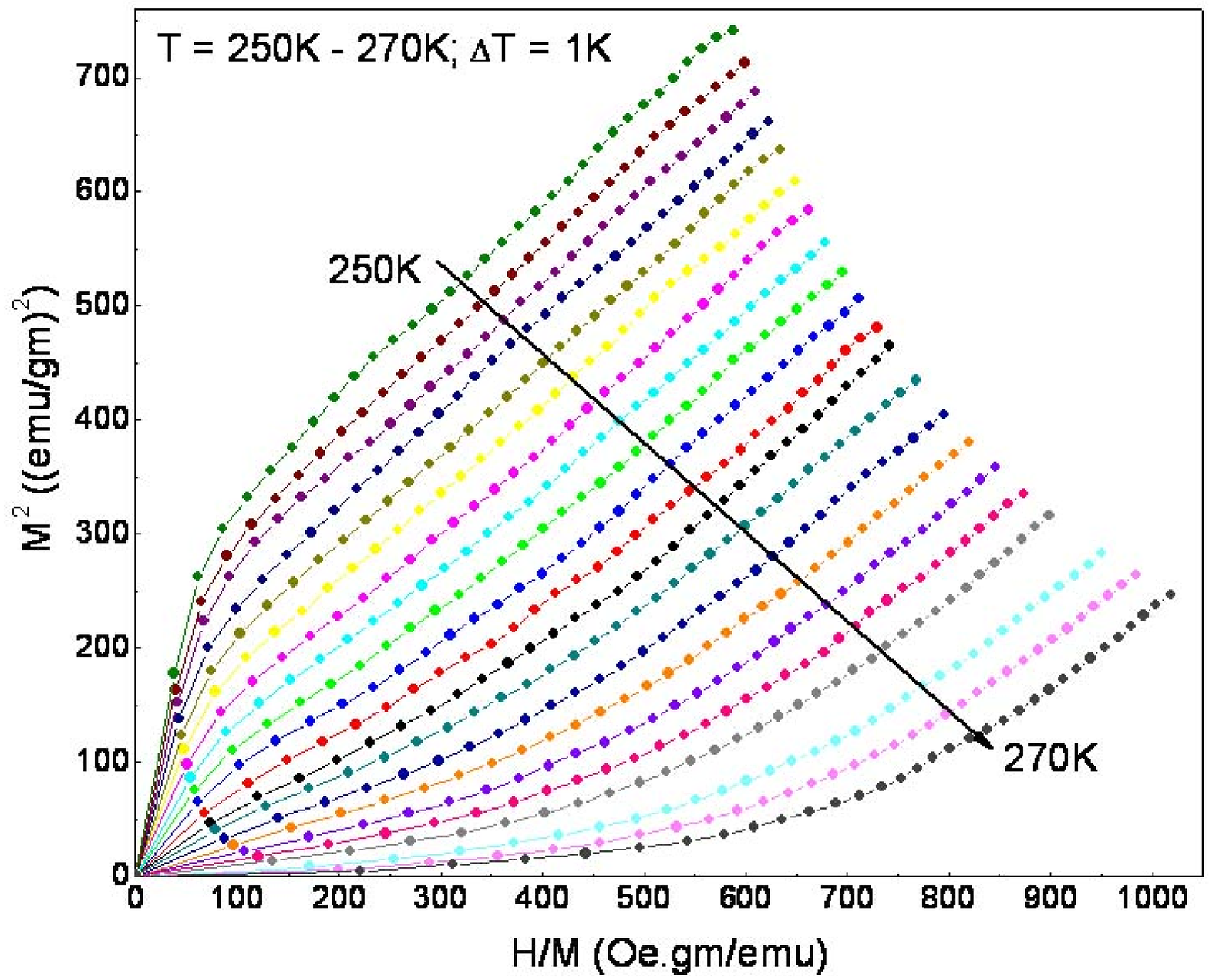}
\end{center}
\caption{Arrott plots for nano LCMO}
\label{Fig5}
\end{figure}

\noindent
The magnetization isotherms for the nanoparticles plotted as Arrott plots are shown in Fig.~\ref{Fig5} taken over a range 250K to 270K. It is seen that the isotherms do not display the anomalous change of slope as seen in the bulk sample. Here, we find a positive slope throughout the range of $M^2$. Nanoparticles of $\mathrm{La_{0.67}Ca_{0.33}MnO_3}$, thus, show a second order magnetic phase transition at $T_C$. The exact values of the critical exponents $\beta$ and $\gamma$ and the exact Curie temperature $T_C$ were determined from a modified Arrott plot by  taking $\beta$, $\gamma$ and  $T_C$ as parameters to be obtained from fit. (The plot  has not been shown to avoid duplication). The exact values of $\beta=0.47\pm 0.01$ and $\gamma=1.06\pm 0.03$ come out to be close to the mean field values ($\beta = 0.5, \gamma = 1$) and  $T_{C}=259K$. The mean field value of the exponent $\delta$ can be derived using $\delta = 1 + \frac{\gamma}{\beta}$. We find $\delta = 3.26 \pm 0.16$. One can also obtain $\delta$ directly from the plot of $M$ vs $H$ at $T_{C}$. From such a plot we obtain $\delta=3.10 \pm 0.13$. A comparison of the critical exponents show that the magnetic transition is not only second order but the exponents are close to the mean field values: $\beta= 0.5$ , $\gamma= 1$  and $\delta= 3$.

\begin{figure}[t]
\begin{center}
\includegraphics[width=8cm,height=7cm]{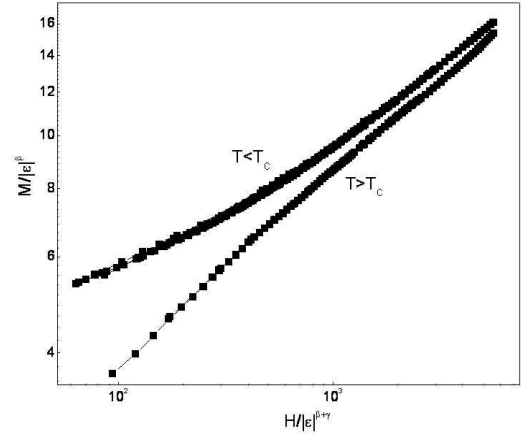}
\end{center}
\caption{Scaling plots for nano LCMO}
\label{Fig6}
\end{figure}

\noindent
The critical exponents of the transition can be equivalently determined by scaling plots of the form $M/|\epsilon|^\beta = f_\pm(H/|\epsilon|^{\gamma+\beta})$, where, $\epsilon = |T-T_C|/T_C$, $f_\pm$ is a scaling function, and the plus and minus sign correspond to the ferromagnetic and paramagnetic regions respectively. By appropriate selection of the parameters $T_C$, $\beta$ and $\gamma$, the data should collapse on two different branches for $T>T_C$ and $T<T_C$. We construct scaling plots (Fig.~\ref{Fig6}) to prove the validity of our choice of $\beta,$ $\gamma$ and $T_C$. A convincing scaling of the data points on the two branches of the scaling function $f_\pm$ can be seen.

\noindent
\section{DISCUSSIONS}

We have observed a very clear change in the nature of the ferromagnetic to paramagnetic phase transition in $\mathrm{La_{0.67}Ca_{0.33}MnO_3}$ from first order to second order when the particle size is brought down from bulk to nano. The nature of paramagnetic transition in double exchange ferromagnet have been investigated theoretically before. The model is an extension of  the compressible spin model of Bean and Rodbell\cite{BR}. It has been found that change over from the first order to second order transition can be parameterized by a quantity $\eta$\cite{ref8} where
\begin{equation}
\eta=K\frac{1}{T_C}(\frac{dT_C}{dP})^2  
\label{eqn1}
\end{equation}

where $K=\frac{35S(S+1)}{6(S-1)(3S+1)}\frac{Nk_{B}}{\kappa T}$, N is the number of magnetic ions /unit volume, $\kappa$ is the compressibility and S is the spin. 

\noindent
The parameter $\eta$ can be obtained experimentally from the $m_{s}-t$ curve where $m_{s}=M_{s}(T)/M_{s}(T=0)$ and $t=T/T_C$. $M_{s}$ is the saturation magnetization. For $\eta <$ 1 the transition is of the second order, while for $\eta \geq 1$  the transition is of the first order. We obtained the value of $\eta$ from the experimental $m_{s}-t$ curves for both the samples using the recipes given in ref 17. We obtain for the bulk sample $\eta$ = 0.75. This agrees very well with the value  of $\eta=0.77$ obtained by past studies. The proximity of $\eta$ to 1 makes the transition first order as expected. Interestingly a similar fit of the $m_{s}-t$ data for the nanoparticles give $\eta \approx 0.03$, which is much less than that seen for the bulk sample and is close to $\eta =0$. The change-over from first order to second order transition is thus reflected in the parameter $\eta$. The important question is what causes this reduction of $\eta$ or the change-over. Below we discuss some of the plausible scenarios that can make this happen.

\noindent
Two features of our results are noteworthy. First, the change in $T_{C}$ is extremely small($\approx 4\%$) in view of the fact that the size changes by nearly 2.5 orders. For itinerant ferromagnets like Ni such size reduction shifts $T_C$ by much larger fraction. We note that in a similar oxide ferromagnetic system like $\mathrm{La_{0.5}Sr_{0.5}CoO_3}$, a similar size reduction (studied by us) changes the $T_{C}$ by 65K (a change of $\sim$ 25\%). Thus the ferromagnetic transition in LCMO is more stable under size reduction. Second, the critical exponents obtained for the transition at $T_{C}$ are very close to the mean field values and the spin system is isotropic.

\noindent
From the experimental value of $\eta = 0.75$  and using S=2, we obtain for the bulk sample $(dT_{C}/dP)_{bulk}\approx 19.9K/GPa $. This agrees well with the value of $dT_{C}/dP \approx 20K/GPa $\cite{ref8} obtained directly for bulk sample at low pressure ($P < 2 GPa$). From the experimental value of $\eta=0.03$ for the nanosample we obtain $(dT_{C}/dP)_{nano}\approx 3.9K/GPa$. This is a significant reduction in the value of $dT_{C}/dP$. We argue that such a reduction in the value of $dT_{C}/dP$ occurs as the effective pressure on the nanoparticles is increased on size reduction. In manganites it has been observed that as the pressure is increased, the $T_{C}-P$ curve reaches a broad peak for pressures close to 4GPa\cite{ref12,ref13,ref14} ($dT_{C}/dP \rightarrow 0$). Beyond that $T_{C}$ decreases slowly as P increases showing that $dT_{C}/dP$ has actually changed sign. It has been argued from structural studies done under pressure\cite{ref14} that at low pressure ($P < 2GPa$) the  enhancement of $T_{C}$ with pressure is due to enhancement of overlap of electron wave functions due to lattice compaction. However, for $P > 3GPa$ there is an enhancement of the Jahn-Teller distortion that halts the increase of $T_{C}$ with $P$. In nanoparticles it is expected that  a surface pressure acts on the particles due to the reduction in the size\cite{JAPPCMO}. From the reduction in the cell volume as observed experimentally we estimated that the nanoparticles, due to size reduction, are under an effective  pressure of the order of 3-4GPa. Assuming the particles to be spherical in shape, we can also estimate the surface pressure $P = 2S/d$, where, d is the diameter of the particle and S is the surface tension. For complex oxides there are uncertainties in the exact value of S but it is in the range of few tens of N/m. Using as a rough estimate $S\approx$ 50N/m\cite{ref9} and an average d = 15nm for our sample, we obtain $P\approx 6$ GPa. This is somewhat larger than but similar to the pressure range at which $T_{C}-P$ curve reaches a peak. It establishes that the size reduction to this range of volume can indeed cause an effective high pressure. Thus a reduction in the value of $dT_{C}/dP$ due to an effective pressure (arising from size reduction)  can be a cause for the reduction in $\eta$ that drives the transition to a second order transition.   

\noindent
Another important effect that can cause the change-over is random strain. The nanoparticles have a hydrostatic pressure caused by the surface pressure as argued before. However, they also have a local random strain. A Williamson-Hall plot\cite{ref6} of the XRD data of the nanocrystals gave a large random strain of $\approx$ 8\%. This random strain can cause rounding off of the first order transition. It is quite likely that the random local strain causes the same physical effect as random substitution because the random substitution due to mismatch of ionic size can also lead to random local strain. Given the susceptibility of the first order transition to random substitution, we feel that this can also be a likely cause. It may happen that the two plausible causes, the spin compressibility under pressure created by size reduction as well as the random strain occur in tandem and cause the change-over.

\noindent
We note that the change-over to second order transition is not a rounding off effect due to size distribution. The sample of  nanoparticles  has a rms size distribution of $\pm$10{\%} estimated from the TEM data  as stated before. On the other hand, the bulk sample has a size distribution of $\pm$14{\%} as estimated from FE-SEM data. If the change-over is due to size distribution we would not have seen a first order transition in the bulk sample. 

\noindent
The nanoparticles can have  a shell of disordered spins at the surface. A simple estimate, taking into account the saturation magnetization values obtained for the bulk and nano LCMO, shows that the disordered surface layer in the nanoparticles of average particle size $\sim$ 15nm has a thickness of $\sim$ 1.2nm. If this shell of disordered spins would have affected the transition then the spin disorder would show up as a relatively large value of  the high field susceptibility as well as absence of a saturation of the magnetization. This does not happen in these samples. Thus the effect of the shell of disordered spins does not seem to have much of an effect on the over all magnetic behavior. 

\noindent
\section{CONCLUSIONS}
In  summary,  we  have shown that in manganites the nature of a phase transition can be changed by tuning the size.  Although bulk $\mathrm{La_{0.67}Ca_{0.33}MnO_3}$ exhibits a first order transition at $T_C$, the nature of the transition can be changed to second order by bringing down the particle size to the nanometer level without any change in the stoichiometry of the samples. The size reduction, however, causes very small ($\sim 4\%$) change in $T_{C}$. This change in the order of the transition is attributed to a lowering of the value of d$T_C$/dP due to an increased pressure in the nanoparticles (surface pressure) and also due to rounding off effect due to random strain. Critical exponents, $\beta, \gamma$ and $\delta$ have also been determined for the nano particle sample. The critical exponents come out close to the mean field values for an isotropic system.

\noindent
\section{ ACKNOWLEDGEMENTS}
The authors would like to thank the Department of Science and Technology, Govt. of India for financial support in the form of a Unit. One of the authors (TS) also thanks UGC, Govt. of India for a fellowship.

\end{document}